\documentclass[manuscript]{aastex}
\usepackage{natbib}
\usepackage{lineno}
\usepackage[figuresright]{rotating}
\usepackage{lscape}
\usepackage{multirow}
\usepackage{rotating}
\usepackage{amsmath}

\shorttitle{Peculiar burst of SAX J1808.4-3658}
\shortauthors{Chen et al.}

\begin{document}


\title{
A faint thermonuclear X-ray burst from SAX~J1808.4--3658 observed with NICER: possible evidence for burst-driven changes in the inner disk structure
}

\author{Yu-Peng Chen\textsuperscript{*}}
\email{chenyp@ihep.ac.cn}
\affil{State Key Laboratory of Particle Astrophysics, Institute of High Energy Physics, Chinese Academy of Sciences, Beijing 100049, China}

\author{Long Ji\textsuperscript{*}}
\email{jilong@mail.sysu.edu.cn}
\affil{School of Physics and Astronomy, Sun Yat-Sen University, Zhuhai, 519082, China}

\author{Shu Zhang\textsuperscript{*}}
\email{szhang@ihep.ac.cn}
\affil{State Key Laboratory of Particle Astrophysics, Institute of High Energy Physics, Chinese Academy of Sciences, Beijing 100049, China}

\author{Shuang-Nan Zhang}
\affil{State Key Laboratory of Particle Astrophysics, Institute of High Energy Physics, Chinese Academy of Sciences, Beijing 100049, China}
\affil{University of Chinese Academy of Sciences, Chinese Academy of Sciences, Beijing 100049, China}

\author{Hua Feng}
\affil{State Key Laboratory of Particle Astrophysics, Institute of High Energy Physics, Chinese Academy of Sciences, Beijing 100049, China}
\affil{University of Chinese Academy of Sciences, Chinese Academy of Sciences, Beijing 100049, China}



\author{Jing-Qiang Peng}
\affil{State Key Laboratory of Particle Astrophysics, Institute of High Energy Physics, Chinese Academy of Sciences, Beijing 100049, China}
\affil{University of Chinese Academy of Sciences, Chinese Academy of Sciences, Beijing 100049, China}

\author{Wen-Zhong Li}
\affil{State Key Laboratory of Particle Astrophysics, Institute of High Energy Physics, Chinese Academy of Sciences, Beijing 100049, China}
\affil{University of Chinese Academy of Sciences, Chinese Academy of Sciences, Beijing 100049, China}

\author{Bo-Yan Chen}
\affil{State Key Laboratory of Particle Astrophysics, Institute of High Energy Physics, Chinese Academy of Sciences, Beijing 100049, China}
\affil{University of Chinese Academy of Sciences, Chinese Academy of Sciences, Beijing 100049, China}



\author{Lian Tao}
\affil{State Key Laboratory of Particle Astrophysics, Institute of High Energy Physics, Chinese Academy of Sciences, Beijing 100049, China}
\affil{University of Chinese Academy of Sciences, Chinese Academy of Sciences, Beijing 100049, China}

\author{Ming-Yu Ge}
\affil{State Key Laboratory of Particle Astrophysics, Institute of High Energy Physics, Chinese Academy of Sciences, Beijing 100049, China}
\affil{University of Chinese Academy of Sciences, Chinese Academy of Sciences, Beijing 100049, China}

\begin{abstract}
We present spectral and temporal properties of a faint thermonuclear X-ray burst observed from SAX~J1808.4--3658 by NICER.
Unlike the previous analysis of the same burst 
\citep{2023ApJ...948..117C}, our analysis focuses on the persistent spectral evolution and changes in the accretion environment.
Following a peak flux of approximately 5\% of the Eddington limit, the burst emission settled into a stable state characterized by a temperature of 0.46 keV and a flux 14\% greater than the pre-burst value in the tail.
Although the observation  was truncated, it resumed after 3 hours, revealing that the source flux returned to its pre-burst level. Our spectral fitting results indicate that at the peak of the burst, the neutron star's surface emission was fully exposed, subsequently becoming increasingly obscured while remaining above the pre-burst emission.
Fitting the persistent spectra also reveals an increase in the visible fraction of the neutron star's surface from 25\% to 41\%  during the period from 100 s to 550 s after the burst, suggesting a change in the disk structure (the height or the inner radius).
The spectral evolution suggests a transient reorganization of the inner accretion flow during the burst peak, followed by partial recovery as the burst flux declines.


\end{abstract}
\keywords{
stars: neutron --- X-rays: individual (SAX~J1808.4--3658) --- X-rays: binaries --- X-rays: bursts}

\section{Introduction}
In a neutron star (NS) X-ray binary (XRB), the NS accretes material from a companion star through an accretion disk or stellar wind, generating X-ray emission.
Meanwhile, the material accumulates on the NS surface.
When the density and pressure at the base of the accreted layer reach critical values, unstable thermonuclear burning is triggered on the NS surface (for reviews, see \citealp{Lewin,Strohmayer,Galloway,2020ApJS..249...32G}), which is named thermonuclear X-ray bursts, also named type I X-ray bursts (hereafter X-ray bursts or bursts).
The luminosity of bursts can reach the Eddington limit and release X-ray emission in 10--100 seconds.
Since bursts occur on the NS, the interplay between bursts and the surrounding accretion circumstances, such as the corona and the accretion disk, has been detected (see, e.g., \citealp{2018SSRv..214...15D}, for a review).

These interactions include, a soft X-ray emission enhancement due to the Poynting effect induced by the burst emission (e.g., \citealp{2013A&A...553A..83I,Worpel2013,Worpel2015,2022ApJ...935..154G});  an iron emission line around 6.4 keV due to the reflection of the burst emission by the disk (e.g., \citealp{2004ApJ...602L.105B,2014ApJ...789..121K,2022MNRAS.509.1736S,2024arXiv240905737S}); the hard X-ray shortage or the  Comptonization/up-scattering of the burst emission by the corona (e.g., \citealp{maccarone2003, Chen2012, Chen2024} and \citealp{Chen16081,Chen16082}).

SAX~J1808.4--3658 (SAX J1808 hereafter) is one of  closest bursters to Earth among the 124 known thermonuclear X-ray bursters\footnote{https://sronpersonalpages.nl/$\sim$jeanz/bursterlist.html}, 
with a distance of approximately 3.5 kpc.
As the first detected accreting millisecond X-ray pulsar (AMXP) with an X-ray pulsation at 401 Hz, an orbital period of around 2.01 hours, and a companion star of 0.043 $M_{\odot}$, which is a Brown Dwarf, SAX J1808 exhibits semiregular outbursts every 2--4 years (e.g., \citealp{2019MNRAS.483..767D}).

Most bursts from SAX J1808.4$-$3658 are photospheric radius expansion (PRE) bursts, indicating peak luminosities up to the Eddington limit. In such bright events, the burst emission can strongly interact with the surrounding accretion environment, producing observable spectral signatures such as additional soft components or disk reprocessing. These effects suggest that thermonuclear bursts are capable of modifying the inner accretion flow, at least in the most luminous cases \citep{2019ApJ...885L...1B,2026MNRAS.548ag624M}.

Whether similar burst--accretion coupling can also be detected in weaker bursts remains much less clear. A faint thermonuclear burst from SAX J1808.4$-$3658 observed with NICER has previously been reported and interpreted in the framework of hydrogen-triggered nuclear burning \citep{2023ApJ...948..117C}. In that picture, unstable hydrogen ignition may either trigger a weak flash or heat the accumulating layer sufficiently to move the envelope toward mixed H/He ignition conditions \citep{2023ApJ...948..117C}. In the present work, we revisit the same event from a different perspective. Rather than focusing primarily on the ignition physics of the burst itself, we examine the evolution of the persistent emission during the burst and investigate whether it carries signatures of a response from the accretion flow.

Our analysis differs from the previous study in two main respects. First, we adopt a different spectral extraction strategy designed to track the burst and persistent components more explicitly in time. Second, we describe the persistent emission using a continuum model more closely tied to the expected accretion components, instead of treating it only as a static background to the burst. This approach allows us to test whether the burst is accompanied by a transient change in the inner accretion environment. We show that the spectral evolution can be interpreted not only in terms of burst-surface emission \citep{2023ApJ...948..117C}, but also in terms of a time-dependent response of the persistent component, possibly associated with burst-driven modification of the inner disk structure. In this sense, our study is complementary to the previous analysis , with particular emphasis on the interplay between the thermonuclear burst and the accretion flow.


 
The paper is organized as follows. the observational data and data reduction procedures of NICER are given in Section 2.
The spectral results and the implications of this thermonuclear burst by NICER are given in Section 3 and Section 4.

\section{Observations and Data Analysis}
Since its launch, NICER has conducted 136 observations (obsids) on SAX J1808, accumulating a total of 820 ks Good Time Intervals (GTIs) between August 13, 2019, and November 5, 2022.
The NICER data were reduced using the pipeline tool nicerl2\footnote{https://heasarc.gsfc.nasa.gov/docs/nicer/nicer\_analysis.html} in NICERDAS 2024-02-09\_V012 with the standard NICER filtering and using ftool nicerl3-lc and nicerl3-spect to extract lightcurves and spectra.
The background is estimated using the 3C50 background model contained in nicerl3-spect command \citep{Remillard2022}.

\begin{figure}[t]
\centering
\includegraphics[angle=0, scale=0.8]{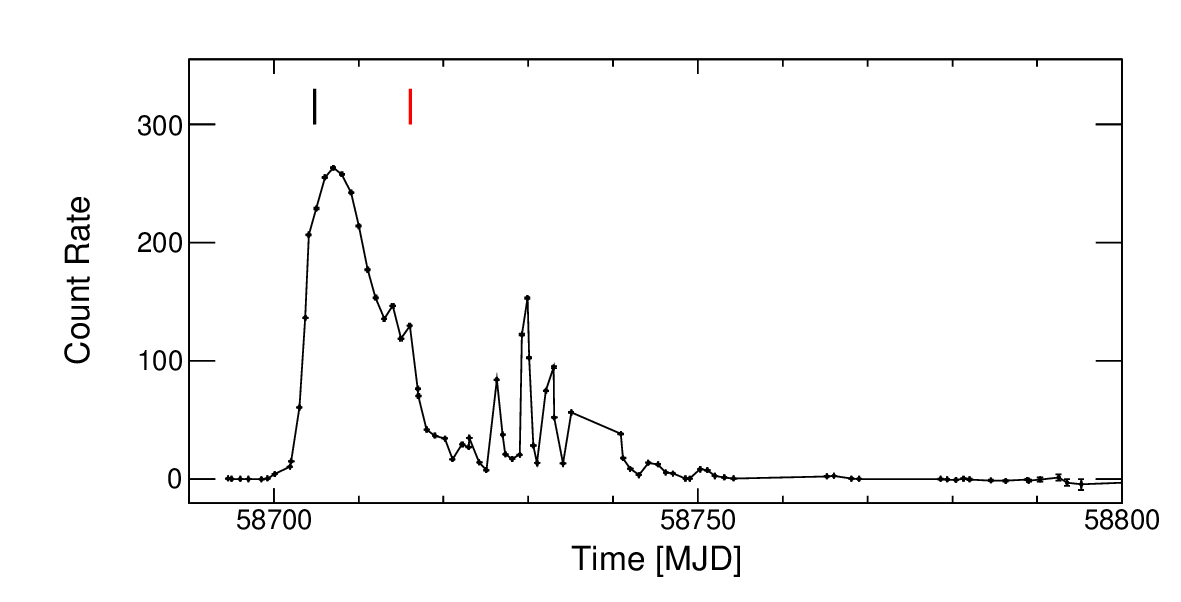} 
 \caption{
Net light curves of SAX J1808.4--3658 by NICER (0.3–12 keV, its full energy band) that are rebinned by one obsid ($\sim$3000 s).
The black and red lines indicate bursts in this work and \citet{2019ApJ...885L...1B}, respectively.
}
\label{fig_outburst_lc}
\end{figure}

We have visually inspected these light curves with a time bin of 1 second. From the light curves, we have identified two bursts with peak fluxes of approximately 1800 cts/s and 33180 cts/s, respectively. These bursts occurred during the rise and decay phases of the outbursts in 2019, as detailed in Table \ref{tb_burst} and illustrated in Figure \ref{fig_outburst_lc}.
The brighter burst has been previously reported by \citet{2019ApJ...885L...1B}, and our focus in this study is on the first and fainter burst.

Additionally, segments of time with elevated background levels, specifically the increased count rates in the 12--15 keV range caused by charged particles, were identified and flagged as suspicious. These periods were excluded from the analysis to maintain the integrity of the data. Light curves were generated in the 12--15 keV energy band with a time bin of 128 seconds, allowing for a detailed examination of the rate distributions. The observational data set (obsid) is organized into five intervals, each with an exposure time ranging from 200 to 1000 seconds. For the five sub-obsids, the count rate in the 12–15 keV range for the burst sub-obsids is 0.05 counts/s, while the count rates for the other sub-obsids range from 0.5 to 1.5 counts/s, which is excessively high for spectral fitting (e.g., \citealp{2023ApJ...946L..24W}); typically, rates of 0.2 to 0.3 counts/s are considered standard. Consequently, only the burst sub-obsid exhibiting bursts was used for spectral analysis.

For burst spectra, since the burst lasts to 500 s after the burst peak, to get the whole evolution, the slices of burst spectra are extracted with variable exposure times.
Specifically, burst spectra are extracted with exposure times ranging from 1 to 40 seconds, ensuring a minimum of 1000 net counts  (pure burst photons 
after subtracting the persistent emission and background) in each spectrum.

The NICER spectra were regrouped using the optimal-binning algorithm in the FTOOL \texttt{ftgrouppha} with a minimum of 25 counts per grouped bin. The systematic error to each channel of the NICER spectrum is added by the ftool niphasyserr.
The equivalent hydrogen column density was 
fixed at $0.21\times 10^{22}~{\rm cm}^{-2}$ \citep{2019MNRAS.483..767D,2026MNRAS.548ag624M} during the spectral fitting.
The resulting spectra are analyzed using XSPEC \citep{Arnaud1996} version 12.14.0. The estimation of the inner disk radius is based on a distance of 3.5 kpc \citep{2020ApJS..249...32G} and an inclination angle of 60$^{\circ}$ \citep{2019MNRAS.483..767D}.

\section{Results}

\subsection{Burst Lightcurve}

\begin{figure}[t]
\centering
\includegraphics[angle=0, scale=0.8]{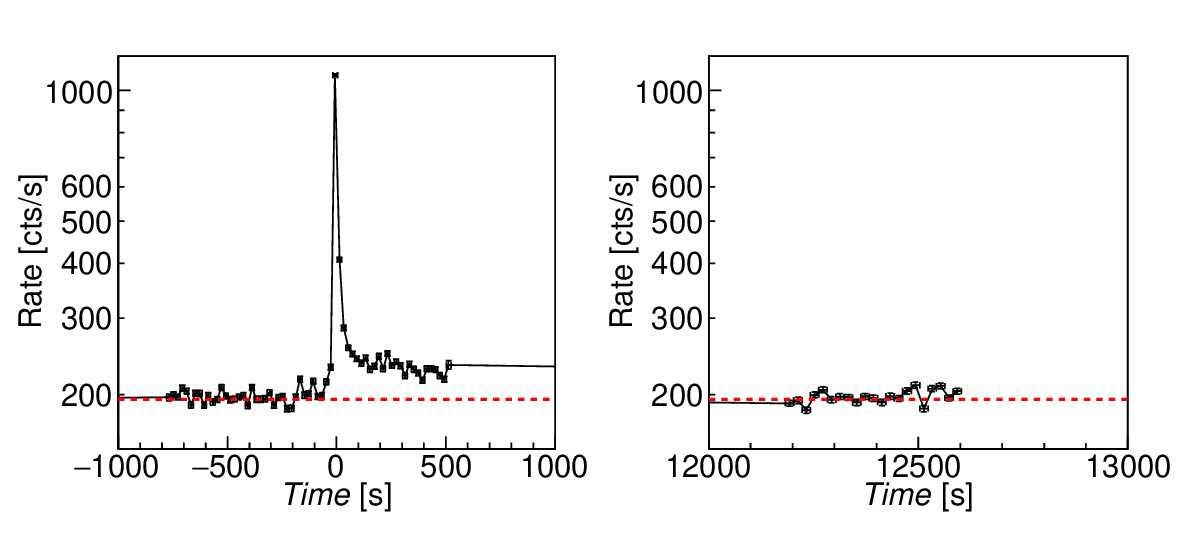} 
 \caption{
Light curves in the NICER observation of SAX~J1808.4--3658 with a time bin 20 s during the burst (left) and after the burst (right) in its full energy band.
  }
\label{fig_burst_lc}
\end{figure}

For the burst, we define the time of the count rate  peak as a time reference (0 point in Figures \ref{fig_fit_burst}. 
We show the burst lightcurves detected by NICER in the left panel of Fig. \ref{fig_burst_lc} with a time resolution of 20 s.
The burst profile exhibits a typical fast rise and slow/exponential decay; especially, the burst has a long tail up to 100 s and then stable with $\sim$40 cts/s above the pre-burst emission lasting for 400 s but the observation was truncated.
After 3.3 hours the observation continued and the flux returned to the pre-burst flux level.
In other words, the increase lasted at least 550 s until the observation was stopped.

\subsection{Pre-burst and post-burst persistent Emission of the burst detected by NICER}

To explore the difference between the emission before and after the burst, the time interval between the start of the sub-obsid and 50 seconds before the burst peak is taken as the time window of the pre-burst emission;
the time interval between 100 seconds after the burst peak and the end of the sub-obsid is taken as the time window of the post-burst emission, i.e., the time intervals are [-800 s, -50 s] and [100 s, 576 s].

We fit the two spectra with the same model.
We choose an absorbed convolution thermal Comptonization model with an input seed photon spectrum to fit the spectra, available as thcomp (a more accurate version of nthcomp) \citep{Zdziarski2020} in XSPEC, which is described by the optical depth $\tau$, the electron temperature $kT_{\rm e}$, and the scattered/covering fraction $f_{\rm sc}$.
The seed photons consist of thermal emission from the accretion disk (diskbb) and/or the  neutron star surface (bbody). 
The initial assumption is that both of thermal components are involved in the Compton scattering, i.e., tbabs*thcomp*(bb+diskbb). 
However, the derived inner disk radius is up to 1000 km, which is non-physical.
In a subsequent analysis, only the disk emission is considered as the seed photons for the corona, modeled by tbabs*(bbodyrad+thcomp*diskbb).
\footnote{We note that our pre-burst persistent spectral model differs from that adopted in the previous analysis by \citet{2023ApJ...948..117C}, whereas the earlier study adopted the more phenomenological model \texttt{phabs*(diskbb+bbodyrad+powerlaw+Gaussian)}. The different baseline spectral descriptions should be kept in mind when comparing the inferred burst spectral evolution and the statistical significance of any additional component. In particular, the inclusion of a simple power-law component may affect the decomposition of the soft X-ray band, where the emission is expected to be dominated mainly by thermal components, and may therefore bias the inferred thermal parameters.
Moreover, we do not include an emission line near 1 keV in our spectral fits, as our primary aim is to model the burst-related continuum evolution rather than narrow spectral features. Previous work by \citet{2023ApJ...948..117C} has shown that excluding this line changes the inferred broadband flux only at the few-percent level and does not significantly affect the main conclusions regarding the source flux.  We therefore adopt the simpler continuum model in the present work.
} 
This model is used to fit the two spectra simultaneously, and the derived Comptonization parameters are found to be consistent with each other. Then the Comptonization parameters for the two spectra are linked, while only the two thermal components are not tied to each other.

\begin{figure}[t]
\centering
\includegraphics[angle=-90, scale=0.35]{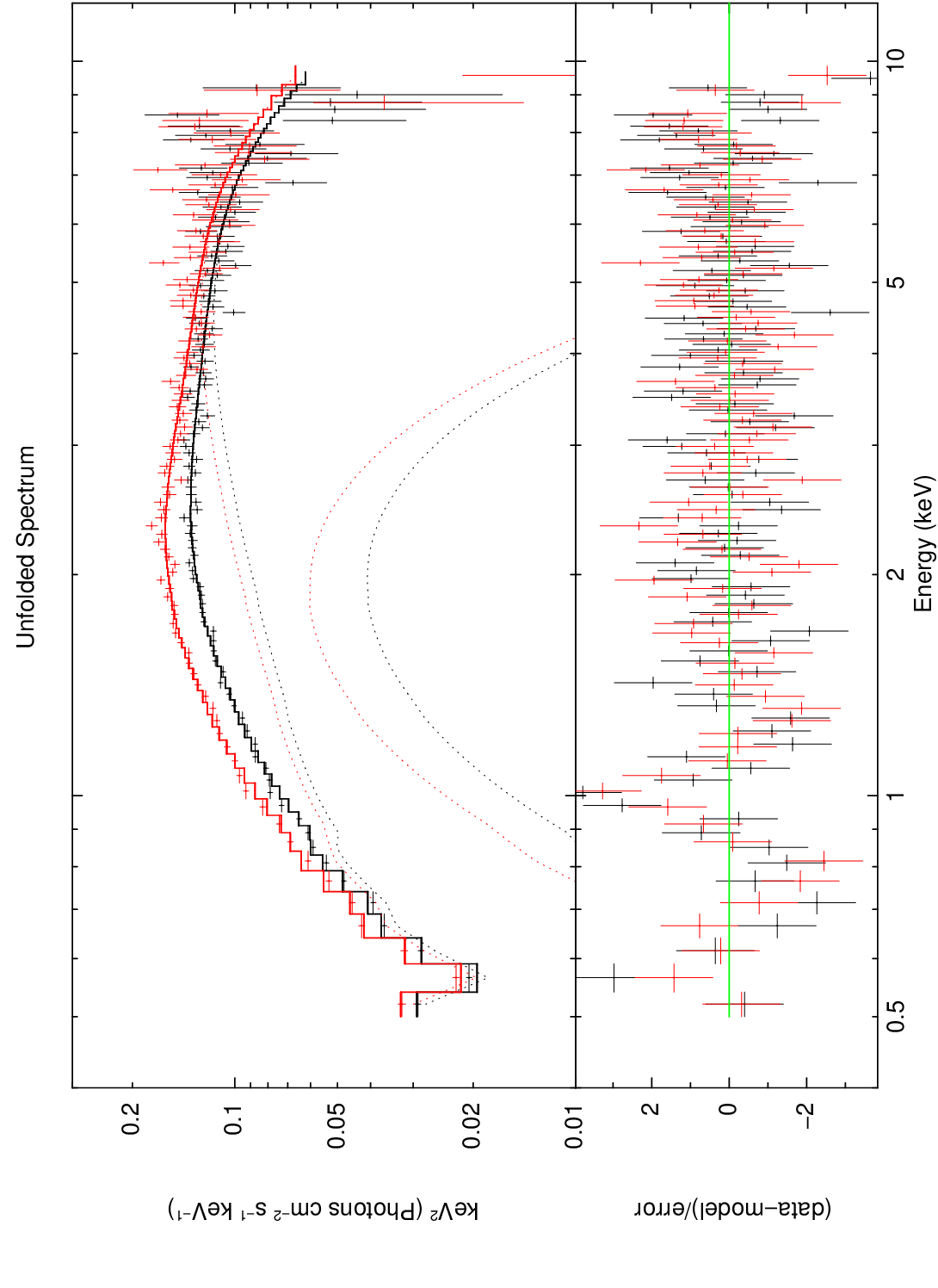}
  \caption{Spectral fitting of before (black) and after (red) the burst by an absorbed blackbody emission and  convolution thermal Comptonization model with an input seed photon spectrum of the accretion disk, i.e., tbabs*(bbodyrad+thcomp*diskbb).
}
\label{fig_outburst_spec}
\end{figure}

We find an acceptable fit, as shown in Table \ref{tb_fit_persist} and Figure \ref{fig_outburst_spec}, with the Comptonization parameters of the electron temperature $kT_{\rm e}$, optical depth $\tau$ and scattered/covering fraction $f_{\rm sc}$ are $1.77^{+0.21}_{-0.18}$ keV, $18.3^{+2.5}_{-4.0}$ and $0.43^{+0.12}_{-0.07}$, respectively.
For the NS surface and the disk emission, the temperatures of the two spectra are approximately the same, but there is an increase in the NS surface radius and the inner disk radius. Specifically, the observed NS surface areas account for 25\% to 41\% of the NS surface for the two spectra, respectively. It is worth noting that all of these areas are less than half of the NS surface area.
Additionally, the post-burst emission is brighter than the pre-burst emission by a factor of 14\%. The bolometric flux $F_{\rm total}$ is measured to be $7.90\pm0.06\times10^{-10}~{\rm erg/cm}^{2}/{\rm s}$ and $F_{\rm total}$ is $9.02_{-0.07}^{+0.05}\times10^{-10}~{\rm erg/cm}^{2}/{\rm s}$, corresponding to 0.64\% $L_{\rm Edd}$ and 0.74\% $L_{\rm Edd}$, respectively. Here, $L_{\rm Edd}$ represents $1.8\times10^{38}~{\rm erg}~{\rm s}^{-1}$, which is the Eddington limit for hydrogen accretion material onto a neutron star with 1.4 solar masses. 


\subsection{Spectral evolution of the burst by the $f_{a}$ model}

Around the burst peak time, taking the pre-burst as the background, the canonical blackbody model was tried to fit the slices of the burst but obtaining a reduced  $\chi^{2}_{\upsilon}\sim$1.4 and with a significant excess at the energy bands of $<2$ keV.
Then a scaling factor is added to multiply the pre-burst emission\footnote{The same burst has been studied previously by \citet{2023ApJ...948..117C}, but with a different spectral extraction strategy, whereas the previous analysis used finer temporal binning with $\sim$500 counts per spectrum. This difference may partly account for the different conclusions regarding the need for an additional spectral component, since our approach emphasizes per-spectrum statistical quality, while the previous one prioritizes temporal resolution.}, i.e., the $f_{a}$ model \citep{Worpel2013}.
In this model, the derived model parameters of the pre-burst emission are fixed and then used to fit the burst spectra (let the blackbody parameters and the constant be variable during spectral fitting), i.e., tbabs*(bbody+fa*(bb+thcomp*diskbb). 
The above model results in reduced  $\chi^{2}_{\upsilon}$=0.8--1.3 with d.o.f. 30--60, as shown in the left panels of Figure \ref{fig_fit_burst}. 
The full fitting results are given in Appendix Table  \ref{tab:combined_fit_results}.


The peak flux of the blackbody is $0.61\pm0.03 \times 10^{-8}~{\rm erg/cm}^2/{\rm s}$. 
The derived NS surface parameters at the burst peak time are 0.97$\pm$0.4 keV and 10.3$\pm0.6$ km.
Additionally, the derived $f_{a}$ reaches its peak 1.8$\pm$0.3 at the burst peak, which corresponds $\sim$23\% peak flux of the burst.

After the burst peak, in the burst decaying phase, the temperature decreases as the the burst decays.
Interestingly, the derived $R_{\rm bb}$ also decreases as the burst decays, which is different with other bursts in previous work.
In current thermonuclear burst theory, $R_{\rm bb}$ should be roughly a constant value in the decaying phase.
In other words, the derived $R_{\rm bb}$ shows a gradual decrease as the burst decays.

In the burst tail phase, e.g., after 100 seconds of the burst peak time, the derived $T_{\rm bb}$ converges to the temperature of the pre-burst--0.47 keV.
Meanwhile, the derived $R_{\rm bb}$ converges to a constant value--roughly 4 km, which is consistent with the difference of NS surface area between the pre-burst emission and post-burst emission.
The derived $f_{a}$ also converges to a constant value--0.05.

We compare the two models mentioned above (the canonical blackbody model and $f_{a}$ model) using the F-test. The results indicate that,  at the burst peak phase, the $f_{a}$ model significantly enhances the fits with a p-value of $<10^{-4}$--$10^{-7}$.
However, at the tail phase, the $f_{a}$ model slightly enhances the fits with a p-value of $<0.1$--$10^{-3}$.

   

Following \citet{Chen2025}, we account for the pre-burst background $F_{\rm pre}$ in the cooling phase analysis. We compare the bolometric flux profiles using power-law (radiative) and exponential (thermal conduction) decay models. As shown in Figure~\ref{fig_fit_burst}, the power-law model provides a significantly better fit ($\chi^2_\nu = 0.7$) than the exponential model ($\chi^2_\nu = 1.7$; 24 d.o.f.). The derived power-law index is $-1.67\pm0.11$, which is steeper than the theoretical value of $-4/3$ for pure radiative cooling \citep{2014A&A...562A..16I}. This steepening is likely driven by the decrease in the derived $R_{\rm bb}$ during the burst decay.


The derived pre-burst NS surface emission, $F_{\mathrm{pre}}$, is estimated to be $(0.37 \pm 0.05) \times 10^{-10}~\mathrm{erg\,cm^{-2}\,s^{-1}}$. This value is consistent with the difference between the pre-burst and post-burst NS surface emissions, i.e.,
\[
(1.5 \pm 0.3) \times 10^{-10} - (1.0 \pm 0.3) \times 10^{-10} = (0.5 \pm 0.4) \times 10^{-10}~\mathrm{erg\,cm^{-2}\,s^{-1}}.
\]

\subsection{Spectral evolution of the burst by the Comptonization model}

Since the burst photons could also be affected by the corona, we investigate whether the model used for the persistent emission could be applied to the burst emission \citep{Chen1816}. In this analysis, the pre-burst emission is no longer considered as background; instead, following the spectral fitting procedure for the persistent emission, the instrumental background of the telescope is utilized as the background emission.

In other words, the model is revised to tbabs*(bbodyrad+thcomp*diskbb) with variable blackbody parameters and fixed Comptonization and accretion disk parameters. However, this approach results in excess at $<$ 2 keV, which is very similar to the residuals by the canonical blackbody model. It is suspected that the model parameters for Comptonization and accretion disk might change during the burst. Then allowing the disk radius to vary leads to a good fit as the $f_{a}$ model, as shown in the right panel of Figure \ref{fig_fit_burst}. 
Subsequently, the model is refined to tbabs*(bbodyrad+thcomp*diskbb) with fixed Comptonization parameters and fixed accretion disk temperature, with the variable parameters being the blackbody temperature and radius, as well as the inner disk radius.

As shown in the right panels of Figure \ref{fig_fit_burst}, the Comptonization model fits the burst well with a reduced $\chi^{2}_{\upsilon}$ as well as the $f_{a}$ model around the burst peak. 
The full fitting results are given in Appendix Table  \ref{tab:combined_fit_results}.
As expected, during the burst peak phase, the blackbody parameters derived from the Comptonization model exhibit consistency within the error bars compared to those obtained using the $f_{a}$ model. Notably, the inner disk radius reaches up to 115$\pm$7 km and decreases in tandem with the burst flux and stables at $\sim$ 70 km at the burst tail phase.

For the NS surface emission, the derived NS surface emission parameters have a similar trend with $f_{a}$ model, e.g., there is also a shortage of the derived $R_{\rm bb}$  as the burst decays.
The radius stabilizes at $\sim$ 6 km, consistent with the values obtained from the post-burst persistent emission.

For the cooling trend, similar to the $f_{a}$ model, the powerlaw model provides an effective description of the burst bolometric flux but not with the exponential function.
The derived $F_{\rm pre}$ $1.4\pm0.04\times10^{-10}~{\rm erg/cm}^2/{\rm s}$ is consistent with the NS surface of the post-burst emission $1.5\pm0.3\times10^{-10}~{\rm erg/cm}^2/{\rm s}$.

To examine how the burst-induced enhancement of the persistent emission is coupled to the burst itself, Figure \ref{fig_fit_fa_bb} shows the enhancement factor $f_a$ as a function of the burst blackbody flux. Two representations are compared. In the left panel $f_a$ is plotted against the directly measured burst flux $F_{\rm bb}$, whereas in the right panel it is plotted against the obscuration-corrected burst flux $F_{\rm bb}^{\rm corr}$, defined as the flux that would be observed if the neutron-star surface emission were not blocked by the inner accretion flow. In both cases $f_a$ increases monotonically with the burst flux and is well described by a power law. For the directly measured flux the relation follows $f_a\propto F_{\rm bb}^{1/2}$, whereas for the corrected flux it steepens to a nearly linear form, $f_a\propto F_{\rm bb}^{\rm corr}$. Because the two panels differ only through the correction for the partial occultation of the neutron-star surface, the change of the power-law index provides a direct, albeit model-dependent, diagnostic of the burst-driven evolution of the inner accretion geometry.

\subsection{Timing analysis}
\begin{figure}
\centering
\includegraphics[width=0.7\linewidth]{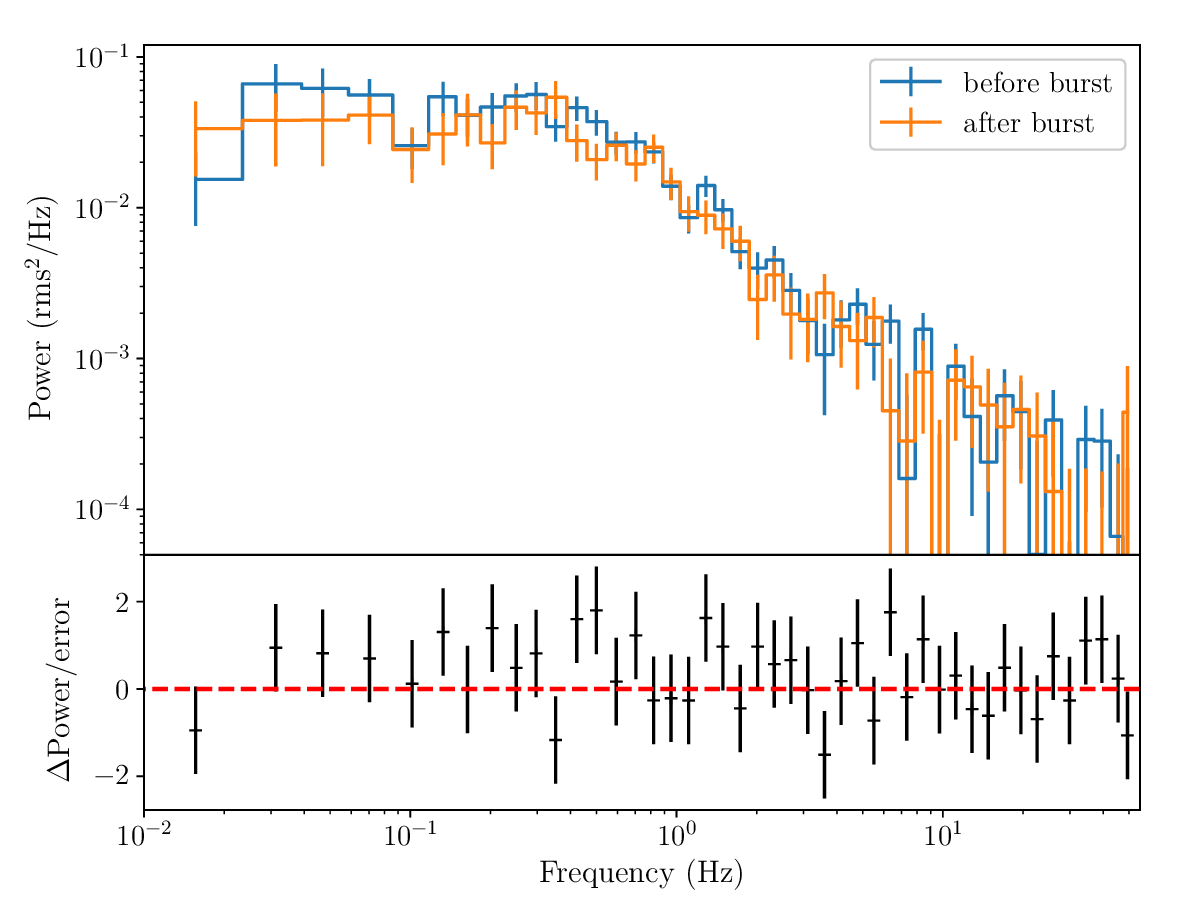}
  \caption{
  The comparison of power spectra obtained before and after the burst and their deviations (bottom panel).
}
\label{compare_powerspec}
\end{figure}

\begin{figure}
\centering
\includegraphics[width=0.7\linewidth]{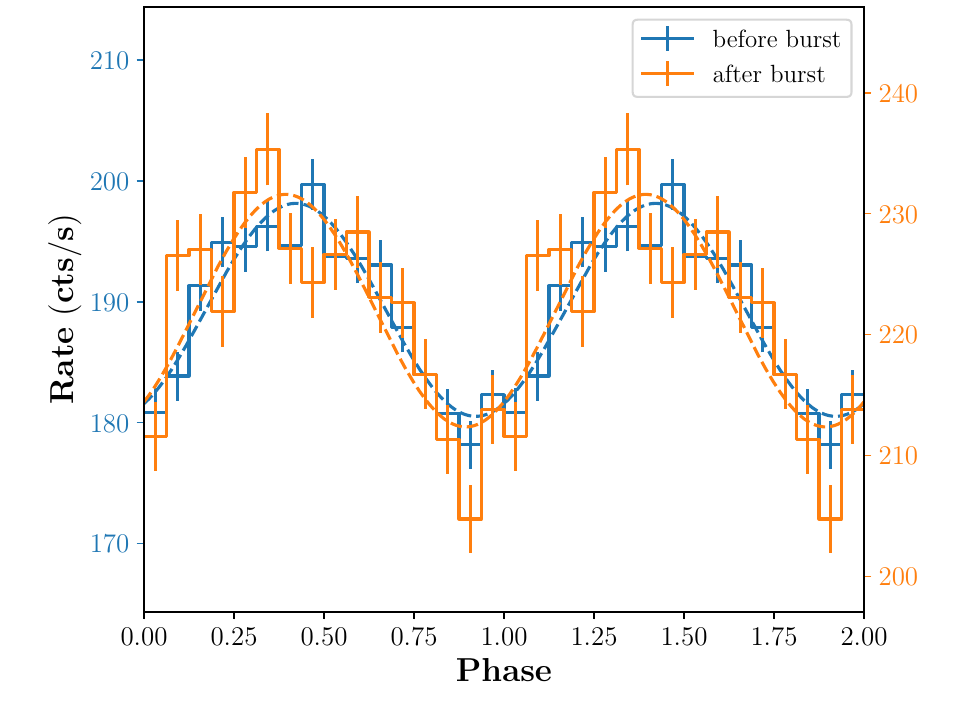}
  \caption{
  The comparison of pulse profiles obtained from the data before and after the burst.
}
\label{compare_pulse_profiels}
\end{figure}

To investigate possible variations induced by the burst, we compared the power spectra and pulsed profiles obtained from the data before and after the burst (Figure~\ref{compare_powerspec} and~\ref{compare_pulse_profiels}).
In Figure~\ref{compare_powerspec}, the root-mean-square (rms) normalization was adopted and the white noise component had been subtracted off.
Generally, we did not find significant changes in power spectra. 
The fractional rms in the frequency range of 0.01--50\,Hz is $23.7_{-0.8}^{+0.4}$\% before the burst and $21.6_{-1.2}^{+0.7}$\% after the burst.
The pulse profiles in Figure~\ref{compare_pulse_profiels} were obtained by folding 0.5--10\,keV lightcurves according to the ephemeris provided by \citet{Illiano2023}.
Generally, they show a similar shape that can be approximately described as a sinusoidal function.
In general, although there may be minor deviations of rms values and the profiles of the persistent emission between the pre-burst and post-burst phases, it is challenging to draw definitive conclusions due to the constraints of limited statistics.

\clearpage

\section{Discussion}

Although the burst studied here is clearly sub-Eddington, our time-resolved spectroscopy shows that it is still capable of perturbing the innermost accretion environment. The burst modifies not only the thermal emission associated with the neutron-star surface, but also the soft X-ray spectral components associated with the inner accretion flow. In this section, we discuss the implications of our results from three related perspectives: the differences between our analysis and previous work on the same event, the physical origin of the burst-induced excess emission and the inferred geometric response of the inner disk, and the constraints that these results place on the recovery of the inner accretion flow. Throughout, we emphasize that the present burst is much weaker than the classical PRE bursts usually invoked in burst--disk interaction studies, and therefore provides a useful test of whether such coupling can also occur in relatively faint thermonuclear bursts.

\subsection{Burst-induced changes in the inner accretion geometry}

Although the burst is faint in an absolute sense, it is still energetically important relative to the persistent emission. The peak burst blackbody flux is $F_{\rm peak}\simeq 0.61\times10^{-8}\ {\rm erg\ cm^{-2}\ s^{-1}}$, corresponding to
\begin{equation}
L_{\rm peak}=4\pi D^2F_{\rm peak}\simeq 9\times10^{36}\ {\rm erg\ s^{-1}},
\end{equation}
for $D=3.5$ kpc. This is about $L_{\rm peak}/L_{\rm Edd}\simeq0.05$, adopting $L_{\rm Edd}=1.8\times10^{38}\ {\rm erg\ s^{-1}}$ for hydrogen-rich material. Thus, the burst is clearly sub-Eddington, and direct radiative expulsion of the accretion disk is unlikely. However, the peak burst luminosity is still far above the pre-burst persistent luminosity, which is only $\sim0.6$--$0.7\%\,L_{\rm Edd}$. From the viewpoint of the innermost accretion flow, the burst therefore represents a strong transient irradiation event.

This is physically relevant because burst--disk coupling does not require an Eddington-level burst. Numerical and observational studies have shown that thermonuclear bursts can perturb the inner accretion flow through a combination of irradiation heating, radiative drag, reprocessing, and structural readjustment of the disk or corona \citep{1989ApJ...346..844W,Worpel2013,Worpel2015,2018SSRv..214...15D}. In particular, Poynting--Robertson drag can remove angular momentum from the innermost disk material and thereby modify the local accretion flow \citep{1989ApJ...346..844W}. In brighter bursts, reflection and burst-induced excess emission have also been used to infer a direct interaction between burst photons and the accretion disk \citep{2004ApJ...602L.105B,2014ApJ...789..121K}.

Our spectral results are consistent with such a transient geometric response. The apparent neutron-star blackbody radius changes from $R_{\rm bb}\simeq4.97$ km before the burst to $R_{\rm bb}\simeq6.37$ km after the burst, while reaching $R_{\rm bb}\simeq10.3$ km near the burst peak. If one adopts $R_{\rm NS}=10$ km, the corresponding visible fractions are
\begin{equation}
f_{\rm vis}=\left(\frac{R_{\rm bb}}{R_{\rm NS}}\right)^2 \simeq 0.25,\ 1.00,\ {\rm and}\ 0.41,
\end{equation}
respectively. Thus, the apparent visible surface area increases by a factor of $\sim4$ from the pre-burst state to the burst peak, and remains larger than the pre-burst value by a factor of $\sim1.6$ afterward. For an inclination of $i=60^\circ$, the unobscured geometrical visible fraction would be
\begin{equation}
f_0=\frac{1+\cos i}{2}\simeq 0.75
\end{equation}
\citep{Shaposhnikov2003}. Since both the pre-burst and post-burst visible fractions remain below this value, the line of sight is likely affected by partial occultation from a vertically extended inner flow. 

A qualitatively similar conclusion is supported by the persistent-emission fits in which the disk component is allowed to vary. In that case, the apparent inner disk radius reaches $\sim115\pm7$ km near the burst peak and later decreases to $\sim70$--75 km in the burst tail and post-burst phases. 
Taken together, these results suggest that the burst temporarily perturbed the inner accretion geometry and reduced the effective obscuration of the neutron-star surface before the disk partially recovered.

\subsection{Interpretation and limitations of the $f_a$ excess}

The statistical preference for the $f_a$ model is strongest near the burst peak, indicating that the burst spectrum cannot be described solely as cooling emission from the neutron-star surface. However, care is required when interpreting this excess physically. Our definition of the $f_a$ component differs from the original variable persistent-emission prescription of \citet{Worpel2013,Worpel2015}. In our analysis, $f_a$ represents the \emph{net increment} of the persistent-like component relative to the pre-burst level, such that $f_a=0$ corresponds to no burst-induced change. 

We therefore argue that the $f_a$ excess is best regarded as a phenomenological proxy for a combination of processes operating simultaneously in the inner accretion environment. These likely include burst irradiation of the inner disk, enhanced disk or boundary-layer reprocessing, time-dependent obscuration of the neutron-star surface, possible redistribution of burst photons through scattering in the inner flow, and perhaps a modest local accretion response. In this sense, the $f_a$ term should not be viewed as a direct measurement of either the disk reprocessing fraction or the instantaneous accretion-rate enhancement. Rather, it captures the part of the burst spectrum that deviates from a pure cooling blackbody and is not explicitly described by the adopted continuum components. 

\subsection{Burst-tail cooling and the shrinking apparent radius}

The burst tail also shows a cooling law that is steeper than expected for pure radiative cooling. The best-fitting power-law index is
\begin{equation}
\alpha=-1.67\pm0.11,
\end{equation}
which is steeper than the canonical $-4/3$ expectation for a purely radiative cooling front \citep{2014A&A...562A..16I}. The exponential model provides a significantly poorer description of the decay. This suggests that the burst tail is not governed by neutron-star surface cooling alone, but also reflects the evolution of the persistent-like component and/or the changing visibility of the neutron-star surface as the accretion geometry evolves.

An important result of this work is that the apparent blackbody radius during the burst cooling tail gradually decreases, rather than following the more common behaviour seen in typical bursts, where the blackbody normalization tends to increase during the early cooling phase in low-luminosity hard/island states, while bursts in high-luminosity soft/banana states often do not show the same trend \citep{2014MNRAS.445.4218K}.  This indicates that the burst evolution cannot be fully described by a simple cooling neutron-star atmosphere with a fixed visible emitting area. A more natural interpretation is that the observed radius is affected not only by surface cooling, but also by changes in the burst viewing geometry and/or the colour-correction factor. If the visible fraction of the neutron-star surface changes during the burst, or if the inner disk partially obscures a time-dependent fraction of the stellar surface, the inferred blackbody radius may evolve even when the true stellar radius remains fixed.

This interpretation is qualitatively consistent with the spectral results of the persistent emission. The visible emitting area inferred from the blackbody component increases strongly during the burst and remains above the pre-burst level afterward, while the apparent inner disk radius evolves significantly during the burst. Taken together, these results support a picture in which the burst temporarily evacuates or vertically modifies the innermost accretion flow, thereby altering the obscuration pattern and the visible emitting area on the neutron-star surface. We therefore suggest that the shrinking apparent radius during the burst tail is not tracing a true contraction of the neutron-star surface, but rather reflects a time-dependent combination of burst cooling, spectral hardening evolution, and partial obscuration by the inner accretion flow. 

\subsection{The $f_a$–$F_{\rm bb}$ relation and the geometry of the inner flow}


The dependence of the burst-induced enhancement of the persistent emission on the burst flux provides a direct probe of how the burst couples to the inner accretion flow. A bright PRE burst from SAX~J1808.4--3658, observed simultaneously with \textit{Chandra} and \textit{RXTE}, also showed a significant burst-induced enhancement of the persistent emission \citep{2013A&A...553A..83I}, with $f_a\propto F_{\rm bb}$. However, \citet{Chen2025} found that the burst-induced enhancement of the accretion flow in 4U~0614+09 scales as $f_a\propto F_{\rm bb}^{1/2}$. Within the Poynting--Robertson (PR) drag framework, in which $L_{\rm PR}=\eta L_{\rm bb}(R_{\rm NS}/R_{\rm d})$, this scaling corresponds to $R_{\rm d}\propto L_{\rm bb}^{1/2}$ and hence to a retreat of the inner disk away from the neutron star.

\begin{figure}[t]
\centering
\includegraphics[angle=0, scale=0.4]{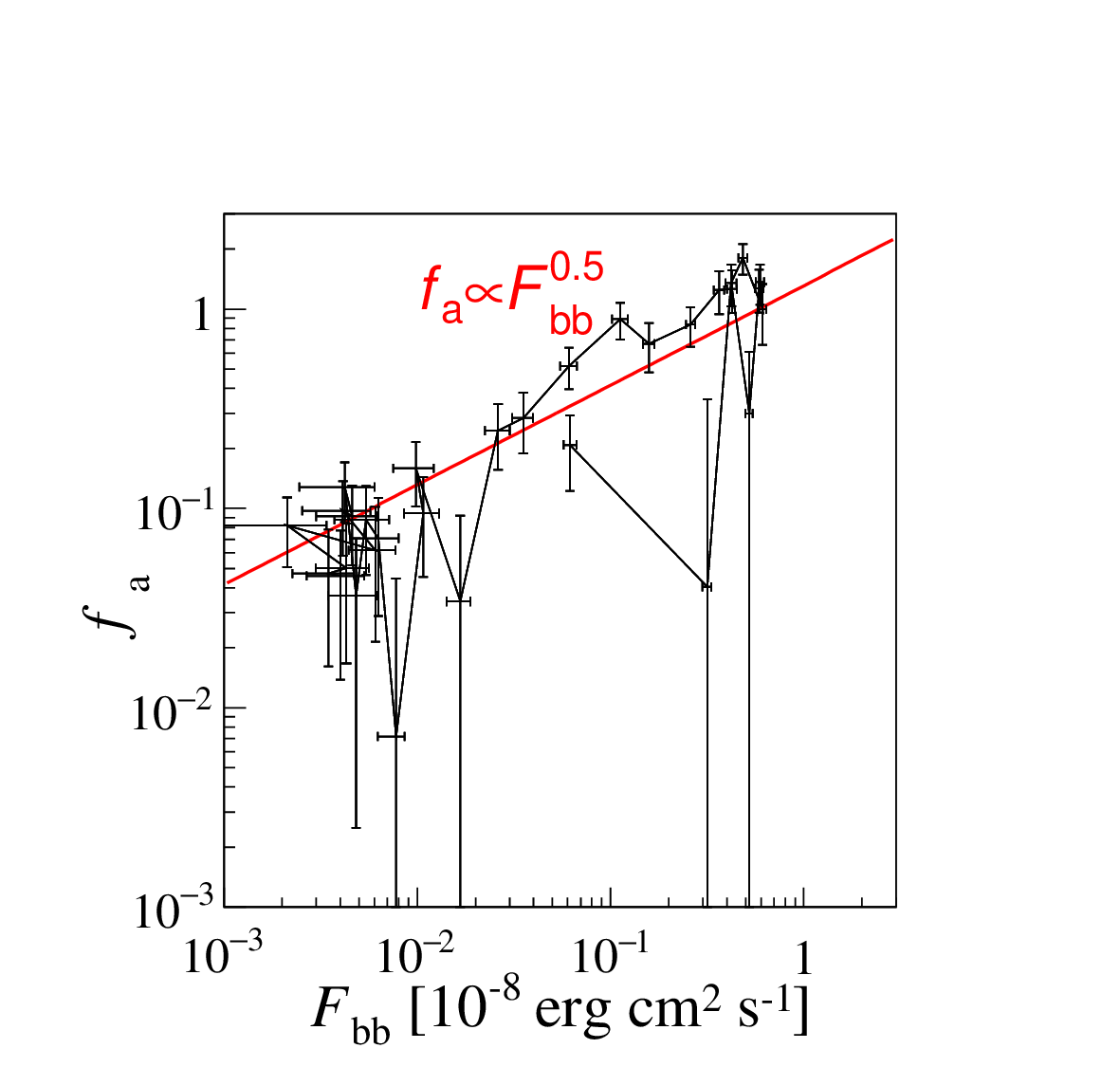}
\includegraphics[angle=0, scale=0.4]{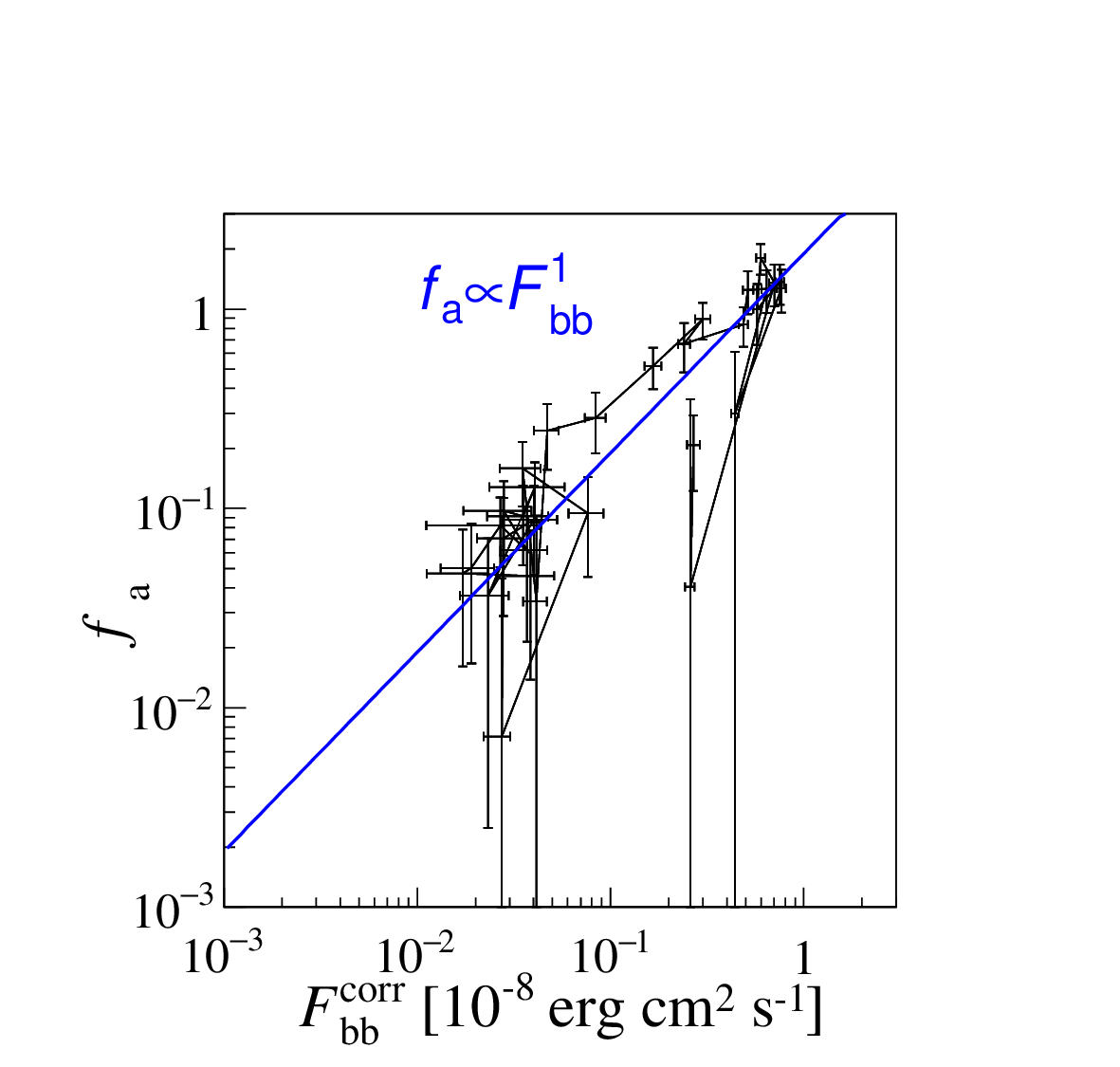}
  \caption{The burst-induced persistent-emission enhancement factor $f_a$ as a function of the burst blackbody flux (in units of $10^{-8}\ \mathrm{erg\ cm^{-2}\ s^{-1}}$). 
  Left: $f_a$ versus the directly measured burst flux $F_{\rm bb}$; the red line is the best-fitting power law $f_a\propto F_{\rm bb}^{1/2}$. 
  Right: $f_a$ versus the obscuration-corrected burst flux $F_{\rm bb}^{\rm corr}$, i.e., the flux that would be received if the neutron-star surface emission were not blocked by the inner accretion flow; the blue line is the best-fitting linear relation $f_a\propto F_{\rm bb}^{\rm corr}$. 
  The steepening of the best-fit index from $\sim1/2$ to $\sim1$ between the two panels reflects the partial occultation of the neutron-star surface, which becomes progressively less severe as the burst brightens.
}
\label{fig_fit_fa_bb}
\end{figure}

For the fainter burst of SAX J1808.4--3658 analysed here, the same 1/2 power law adequately describes $f_a$ when the directly measured burst flux is used (Figure \ref{fig_fit_fa_bb}, left panel). However, when the partial occultation of the neutron-star surface is taken into account, i.e., when $f_a$ is plotted against the obscuration-corrected flux $F_{\rm bb}^{\rm corr}$, the relation steepens to a nearly linear form, $f_a\propto F_{\rm bb}^{\rm corr}$ (Figure \ref{fig_fit_fa_bb}, right panel). This change of index has a natural geometrical origin, and it is closely linked to the changing visibility of the neutron-star surface discussed in Section 4.3.

The same inner-disk geometry that controls the occultation also governs the efficiency of the PR drag. In the general expression for the burst-induced enhancement \citep{1989ApJ...346..844W,Chen2025}:

\begin{equation}
\begin{split}
    L_{\mathrm{PR}} &= \eta g L_{\mathrm{bb}} \\
    &= \eta L_{\mathrm{bb}} \max \left( \frac{R_{\mathrm{NS}}}{R_{\mathrm{disk}}}, \, \frac{\mathrm{d}H}{\mathrm{d}R_{\mathrm{disk}}} \right)
\end{split}
\end{equation}
here $\eta = GM/R_{\rm NS}c^{2} \simeq 0.2$ for a NS X-ray binary, $L_{\rm bb}$ is the burst luminosity, $R_{\rm NS}$ is the NS radius, $R_{\rm disk}$ is the inner disk radius, and $H$ is the height of the inner disk.
Only in the simplified thin-disk limit, with a fixed scale-height gradient, does the geometric factor $g$ reduce to $R_{\rm NS}/R_{\rm d}$, for which a retreating disk gives $L_{\rm PR}\propto L_{\rm bb}^{1/2}$. If instead the disk opening angle grows as the inner radius recedes, as expected for an irradiated, flaring disk in which $H/R$ increases with $R$, the decline of $R_{\rm NS}/R_{\rm d}$ is partially compensated, and the effective index steepens from 1/2 towards unity. The nearly linear corrected relation found here therefore does not require the inner disk to remain stationary; it can equally be produced by a disk that recedes while its scale height grows, so that the burst simultaneously enhances the accretion flow through PR drag and reduces its own occultation of the neutron-star surface.

\subsection{Recovery timescale and implications for the inner flow}

The post-burst observation indicates that the source flux returned to its pre-burst level only after the burst had ended and the observation resumed several hours later. The recovery timescale is therefore constrained to be between $\sim550$ s and $\sim1.2\times10^4$ s. This is broadly consistent with a viscous refill time for the inner disk \citep{1973A&A....24..337S},
\begin{equation}
t_{\rm visc}\simeq \frac{1}{\alpha}\left(\frac{R}{H}\right)^2\Omega_{\rm K}^{-1},
\end{equation}
where $\Omega_{\rm K}=(GM/R^3)^{1/2}$ is the Keplerian frequency. 
For a thin inner disk at $R\sim70$--100 km, the observed recovery timescale can be accommodated by standard thin-disk parameters, particularly for relatively small values of $\alpha$ and $H/R$ \citep{1973A&A....24..337S}.
Specifically, adopting relatively small values of $\alpha \sim 0.01$ and $H/R \sim 0.01$ yields a viscous timescale of $t_{\rm visc} \sim 3100$~s, which falls within the observed range. 
Since the burst duration itself is only of order $10^2$~s, it is too short for complete viscous recovery. Therefore, partial depletion or vertical restructuring of the inner disk during the burst is physically plausible.

It is worth emphasizing that the post-burst continuum enhancement described
above would have been extremely difficult to detect with previous X-ray
missions, which helps explain why this effect has not been widely reported.
The soft blackbody component associated with the enhanced accretion flow has
a characteristic temperature of $kT_{\rm bb} \sim 0.45$~keV, placing the
bulk of its emission well below the low-energy threshold of the RXTE/PCA
($\sim 3$~keV).  Consequently, the RXTE/PCA (5 PCUs) could only detect a modest excess of $\sim 5$~cts~s$^{-1}$ above the persistent background in its energy band, making the spectral signature marginal and highly susceptible to contamination from the underlying persistent spectrum. 
Similarly, X-ray instruments such as NuSTAR (FPM-A \& FPM-B), Swift/XRT, Insight-HXMT/LE, and EP/FXT (FXT-A \& FXT-B, \citealp{2018SPIE10699E..25Y}) predict count rates of only $\sim 0.4$, $1.7$, 
$1.7$ and $8$~cts~s$^{-1}$, respectively, from the enhanced blackbody---well
below the level needed for a clean spectral decomposition. 
The future eXTP/SFA mission \citep{2025SCPMA..6819502Z} , with its large effective area in the soft X-ray band, would measure a count rate of $\sim 80$~cts~s$^{-1}$ from the same
enhanced component enabling a high-significance detection and robust spectral decomposition.

The detectability also depends on the distance to the source.  SAX~J1808.4$-$3658 lies at a relatively nearby distance of $\sim 3.5$~kpc, which enhances the
observed flux.  For more distant atoll and Z sources, the effect would be
further dimmed.  For example, 4U~1636$-$536, at a distance of $\sim 6$~kpc, the enhanced blackbody
flux would be reduced by a factor of $(6/3.5)^2 \approx 3$ in flux, yielding
a NICER count rate of only $\sim 13$~cts~s$^{-1}$.
In sources beyond $\sim 10$~kpc,
even eXTP/SFA would face significant challenges.  These considerations
highlight that the detection of post-burst continuum enhancement is
inherently biased toward nearby, low-accretion-rate systems, and that
 high-throughput soft X-ray monitors like eXTP/SFA are essential
for building a complete census of this phenomenon across the full population
of low-mass X-ray binaries.



\section{Summary and Conclusions}

We have revisited the faint thermonuclear X-ray burst from SAX J1808.4$-$3658 observed with \textit{NICER}, with particular emphasis on the time-dependent response of the accretion flow to the burst. Unlike previous work that focused mainly on the ignition physics of this event \citep{2023ApJ...948..117C}, our analysis explores whether even a sub-Eddington burst can leave measurable signatures in the innermost accretion environment. Our main results can be summarized as follows.

\begin{enumerate}

\item The burst,  with a peak luminosity about $0.05\,L_{\rm Edd}$,  is not fully described by neutron-star surface emission alone. Near the burst peak, the spectra show a statistically significant excess persistent-like component, indicating that an additional soft spectral contribution is required beyond a simple cooling blackbody. 

\item The apparent neutron-star emitting area changes substantially during the burst. The inferred blackbody radius increases from $\sim 5$ km before the burst to $\sim 10$ km near the burst peak, and then settles at a post-burst value of $\sim 6.4$ km. 

\item The apparent inner disk radius also evolves significantly during the burst. In our fits, the radius reaches $\sim 115$ km near the burst peak and later decreases to $\sim 70$--75 km in the burst tail and post-burst phases.

\item A notable feature of this burst is that the apparent blackbody radius gradually decreases during the cooling tail, 
which differs from the behavior observed in previous bursts.


\end{enumerate}

Overall, our results support the view that even a relatively faint, sub-Eddington thermonuclear burst can perturb the innermost accretion environment in SAX J1808.4$-$3658. The burst appears to modify the soft X-ray emitting region, alter the visible fraction of the neutron-star surface, and temporarily perturb the inner disk structure, with the post-burst evolution being broadly consistent with a viscous recovery of the inner flow. 

Future joint observations with Insight-HXMT, EP and the future eXTP will be crucial. Their simultaneous soft and hard X-ray coverage will provide a much cleaner separation between burst emission and accretion-flow response. Such data should allow direct tests of disk reprocessing, inner-disk disruption or recession, and coronal cooling during weak thermonuclear bursts.

\clearpage

\acknowledgements
This research has made use of data and software provided by of data obtained from the High Energy Astrophysics Science Archive Research Center (HEASARC), provided by NASA’s Goddard Space Flight Center.
This work is supported by 
the China's Space Origins Exploration Program, 
the National Key R\&D Program of China (2021YFA0718500) 
and the National Natural Science Foundation of China under grants No. 12673056 and 12261141691.

\clearpage
\bibliographystyle{plainnat}

\clearpage


\begin{table}[ptbptbptb]\small
\begin{center}
\caption{The burst parameters  of SAX~J1808.4--3658 detected by NICER in this work: obsid, peak time $t_{\rm p}$, fluence   $E_{\rm b}^{bb}$,  peak flux $F_{\rm p}^{bb}$   derived from the $f_{\rm a}$ model, burst duration $\tau$ from  $E_{\rm b}^{bb}$/$F_{\rm p}^{bb}$.
}
\label{tb_burst}
\begin{tabular}{cccccccccc}
\\\hline
 obsid & $t_{\rm p}$  & $E_{\rm b}^{bb}$ & $F_{\rm p}^{bb}$  & $\tau$   \\\hline
     	&  MJD   & $10^{-8}~{\rm erg}~{\rm cm}^{-2}$  &  $10^{-8}~{\rm erg}~{\rm cm}^{-2}~{\rm s}^{-1}$   & s    \\\hline
  2584010102 & 58704.8077 & $7.18\pm0.02$ & $0.61\pm0.03$   & $11.8\pm0.6$ 
  \\\hline
\end{tabular}
\end{center}
\end{table}

\begin{table}\small
\centering
\caption{Spectral fitting results of the persistent emission before and after the burst, modeled with tbabs*(bbodyrad+thcomp*diskbb) in the 0.5–10 keV energy band.
}
\label{tb_fit_persist}
\begin{tabular}{ccccccccc}
\\\hline
 State & $kT_{\rm bb}$ & $R_{\rm bb}$&$\tau$ & $kT_{\rm e}$  & $f_{\rm sc}$ & $kT_{\rm disk}$ & $R_{\rm disk}$ & $F_{\rm total}$ \\\hline
Pre-burst & $0.47\pm0.02$ &$4.97_{-0.75}^{+1.45}$ & $18.3_{-4.0}^{+2.5}$ & $1.77_{-0.18}^{+0.21}$ & $0.43_{-0.07}^{+0.12}$ & $0.18_{-0.01}^{+0.02}$ & $65.9_{-10.9}^{+20.4}$ & $7.90\pm0.06$ \\\hline
Post-burst & $0.46_{-0.01}^{+0.02}$ & $6.37_{-0.81}^{+1.12}$ & - & - & - & $0.17\pm0.02$ & $73.9_{-12.8}^{+25.7}$ & $9.02_{-0.07}^{+0.05}$ \\\hline
\end{tabular}
\begin{list}{}{}
\item[a] The model parameters: 
the optical depth $\tau$,  the  electron temperature $kT_{\rm e}$,  the cover factor $f_{\rm sc}$;  the accretion disk  temperature   $kT_{\rm disk}$ and  the inner disk radius   $R_{\rm disk}$ at a distance of 3.5 kpc and an inclination angle of $\theta$=60$^{\circ}$ \citep{2019MNRAS.483..767D}, the total  bolometric flux  $F_{\rm total}$ is in units of $10^{-9}~{\rm erg/cm}^{2}/{\rm s}$.
\end{list}
\end{table}
 
\clearpage

\begin{figure}[t]
\centering
\includegraphics[angle=0, scale=0.4]{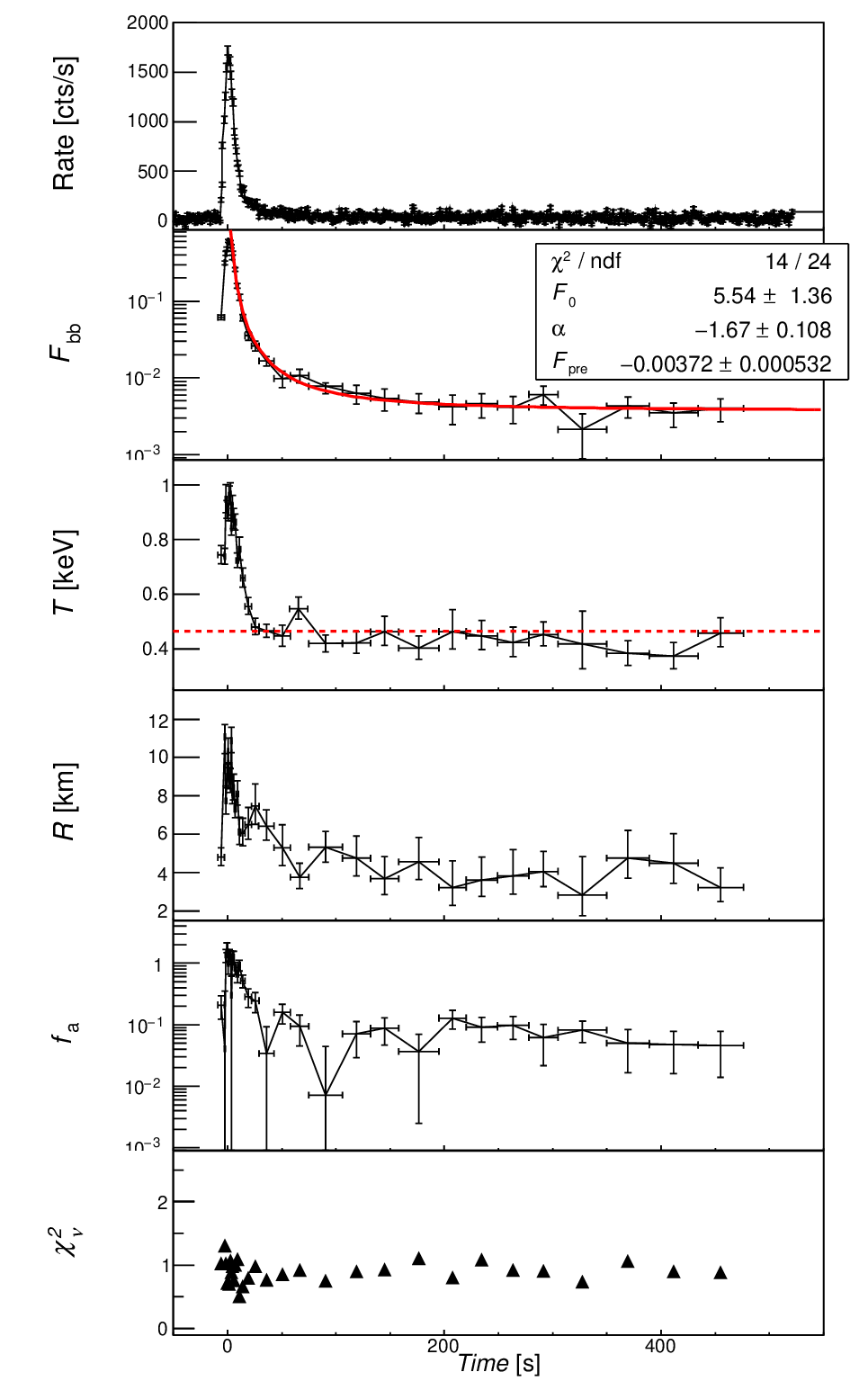}
\includegraphics[angle=0, scale=0.4]{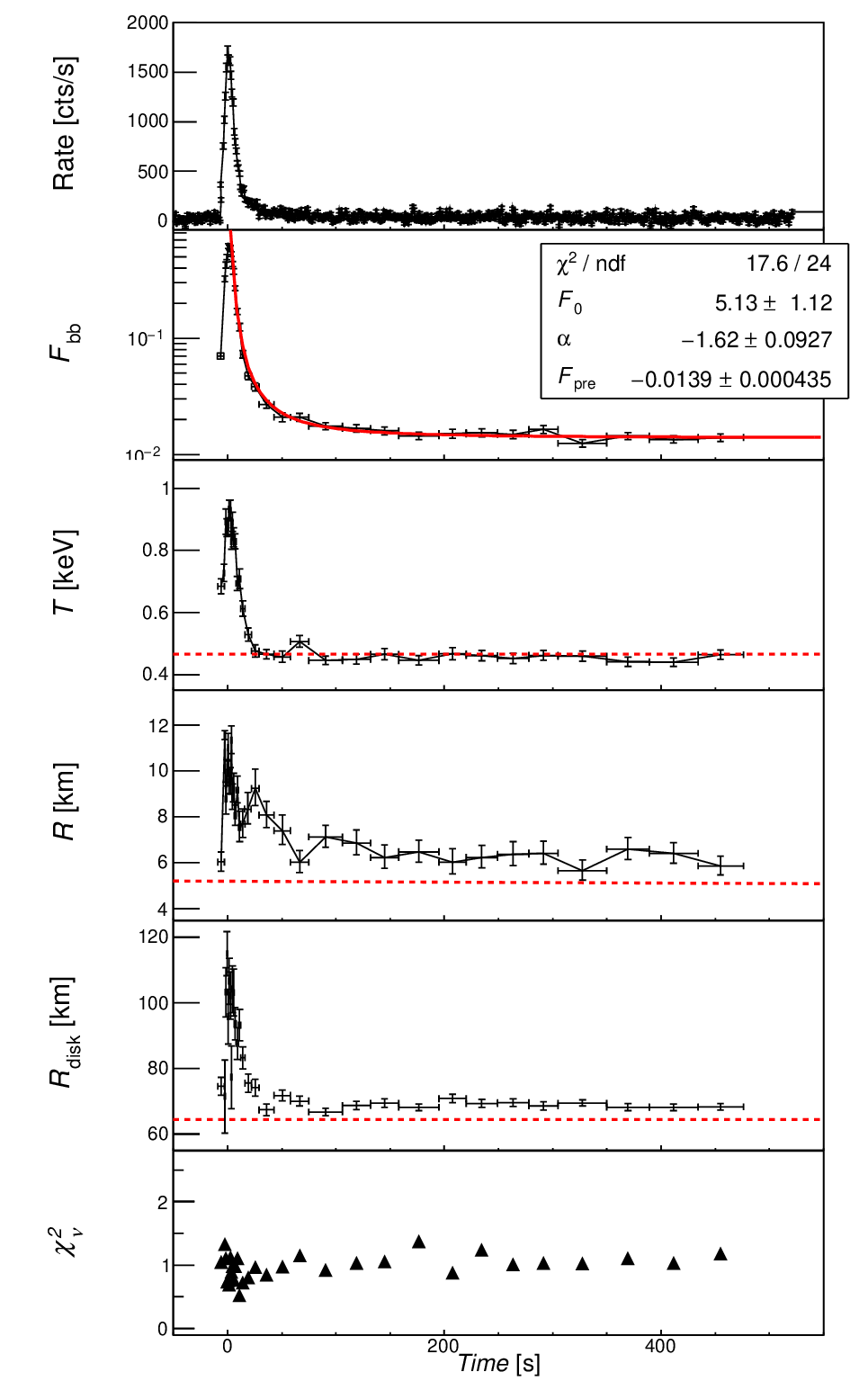}
  \caption{
The spectral fitting results of the burst using the $f_{a}$ model (left panel) and convolution thermal-Comptonization model (right panel).
This includes the time evolution of the blackbody bolometric flux $F_{\rm bb}$, the temperature $kT_{\rm bb}$, the observed radius $R$ of the NS surface at 3.5 kpc, $f_{a}$, the inner disk radius $R_{\rm disk}$, and the goodness of fit $\chi_{v}^{2}$. The bolometric flux of the blackbody model $F_{\rm bb}$ is in units of $10^{-8}~{\rm erg/cm}^{2}/{\rm s}$. 
The red lines  indicate the fitting results of the time evolution of $F_{\rm bb}$ at the decay phase using the revised power law function, i.e., $F_{\rm bb}=F_{\rm 0}t^{\alpha}-F_{\rm pre}$.
The red dashed lines in the evolution of $kT_{\rm bb}$, $R$, and $R_{\rm disk}$ indicate the pre-burst values of the temperature, the radius, and the inner disk radius, respectively.
}
\label{fig_fit_burst}
\end{figure}

\clearpage
\appendix
\setcounter{table}{0}
\renewcommand{\thetable}{A\arabic{table}}
\section{The results of spectral fitting with different models}

\begin{sidewaystable*}
\tiny
\centering
\caption{Time-resolved spectral fitting results of the burst with the blackbody model, the $f_{\rm a}$ model, and the thermal-Comptonization model in 0.5--10 keV. The last two columns give the F-test p-values comparing the BB model with the $f_{\rm a}$ model and the thermal-Comptonization model, respectively.}
\label{tab:combined_fit_results}
\setlength{\tabcolsep}{3pt}
\begin{tabular}{c|cccc|ccccc|ccccc|cc}
\hline
\multirow{2}{*}{Time (s)} 
& \multicolumn{4}{c|}{BB model} 
& \multicolumn{5}{c|}{$f_{\rm a}$ model} 
& \multicolumn{5}{c|}{Thermal-Comptonization model}
& \multicolumn{2}{c}{F-test p-value} \\
\cline{2-17}
& $kT_{\rm bb}$ & $R_{\rm bb}$ & $F_{\rm bb}$ & $\chi_\nu^2$
& $kT_{\rm bb}$ & $R_{\rm bb}$ & $F_{\rm bb}$ & $f_{\rm a}$ & $\chi_\nu^2$
& $kT_{\rm bb}$ & $R_{\rm bb}$ & $F_{\rm bb}$ & $R_{\rm disk}$ & $\chi_\nu^2$
& BB vs $f_{\rm a}$ & BB vs TC \\
\hline
$-6.0\pm3.0$   & $0.71_{-0.02}^{+0.03}$ & $5.6_{-0.3}^{+0.3}$ & $0.069_{-0.003}^{+0.003}$ & 1.15(47) & $0.74_{-0.03}^{+0.03}$ & $4.8_{-0.4}^{+0.5}$ & $0.062_{-0.005}^{+0.005}$ & $0.21_{-0.09}^{+0.09}$ & 1.05(46) & $0.68_{-0.02}^{+0.02}$ & $6.0_{-0.4}^{+0.4}$ & $0.070_{-0.004}^{+0.004}$ & $74.6_{-2.7}^{+2.7}$ & 1.07(46) & $2.4\times10^{-2}$ & $3.6\times10^{-2}$ \\
$-2.5\pm0.5$   & $0.73_{-0.03}^{+0.03}$ & $11.2_{-0.6}^{+0.6}$ & $0.318_{-0.016}^{+0.016}$ & 1.31(34) & $0.74_{-0.03}^{+0.03}$ & $11.1_{-0.9}^{+0.6}$ & $0.317_{-0.019}^{+0.015}$ & $0.04_{-0.04}^{+0.31}$ & 1.35(33) & $0.72_{-0.03}^{+0.03}$ & $11.6_{-0.2}^{+0.2}$ & $0.324_{-0.018}^{+0.018}$ & $71.5_{-11.2}^{+11.0}$ & 1.37(33) & $9.0\times10^{-1}$ & -- \\
$-1.5\pm0.5$   & $0.84_{-0.03}^{+0.03}$ & $10.2_{-0.5}^{+0.6}$ & $0.444_{-0.023}^{+0.024}$ & 1.45(40) & $0.95_{-0.05}^{+0.05}$ & $7.7_{-0.7}^{+0.8}$ & $0.421_{-0.027}^{+0.028}$ & $1.35_{-0.32}^{+0.32}$ & 1.05(39) & $0.89_{-0.04}^{+0.04}$ & $8.8_{-0.6}^{+0.7}$ & $0.425_{-0.025}^{+0.026}$ & $103.3_{-7.5}^{+7.4}$ & 1.14(39) & $2.7\times10^{-4}$ & $1.4\times10^{-3}$ \\
$-0.5\pm0.5$   & $0.83_{-0.02}^{+0.02}$ & $11.5_{-0.5}^{+0.5}$ & $0.545_{-0.022}^{+0.023}$ & 1.40(45) & $0.91_{-0.03}^{+0.03}$ & $9.0_{-0.6}^{+0.6}$ & $0.484_{-0.026}^{+0.026}$ & $1.81_{-0.32}^{+0.32}$ & 0.73(44) & $0.87_{-0.03}^{+0.03}$ & $9.9_{-0.5}^{+0.6}$ & $0.496_{-0.025}^{+0.025}$ & $115.0_{-6.7}^{+6.7}$ & 0.75(44) & $7.5\times10^{-8}$ & $1.3\times10^{-7}$ \\
$0.5\pm0.5$    & $0.85_{-0.02}^{+0.02}$ & $11.8_{-0.5}^{+0.5}$ & $0.637_{-0.025}^{+0.026}$ & 0.91(47) & $0.90_{-0.03}^{+0.03}$ & $10.3_{-0.6}^{+0.7}$ & $0.611_{-0.028}^{+0.028}$ & $1.00_{-0.34}^{+0.33}$ & 0.75(46) & $0.87_{-0.03}^{+0.03}$ & $11.0_{-0.6}^{+0.6}$ & $0.618_{-0.027}^{+0.027}$ & $95.8_{-8.4}^{+8.3}$ & 0.78(46) & $1.4\times10^{-3}$ & $4.2\times10^{-3}$ \\
$1.5\pm0.5$    & $0.89_{-0.02}^{+0.02}$ & $10.7_{-0.4}^{+0.4}$ & $0.634_{-0.026}^{+0.027}$ & 1.09(48) & $0.96_{-0.03}^{+0.03}$ & $8.9_{-0.5}^{+0.6}$ & $0.594_{-0.029}^{+0.030}$ & $1.36_{-0.31}^{+0.31}$ & 0.72(47) & $0.93_{-0.03}^{+0.03}$ & $9.5_{-0.5}^{+0.5}$ & $0.600_{-0.028}^{+0.028}$ & $106.6_{-7.0}^{+7.0}$ & 0.71(47) & $5.3\times10^{-6}$ & $4.1\times10^{-6}$ \\
$2.5\pm0.5$    & $0.89_{-0.03}^{+0.03}$ & $10.7_{-0.5}^{+0.5}$ & $0.612_{-0.027}^{+0.028}$ & 1.42(47) & $0.97_{-0.04}^{+0.04}$ & $8.8_{-0.6}^{+0.7}$ & $0.588_{-0.030}^{+0.031}$ & $1.27_{-0.31}^{+0.31}$ & 1.10(46) & $0.93_{-0.03}^{+0.03}$ & $9.5_{-0.6}^{+0.6}$ & $0.591_{-0.029}^{+0.030}$ & $102.2_{-7.2}^{+7.2}$ & 1.15(46) & $3.6\times10^{-4}$ & $1.1\times10^{-3}$ \\
$3.5\pm0.5$    & $0.83_{-0.02}^{+0.02}$ & $11.3_{-0.5}^{+0.5}$ & $0.531_{-0.021}^{+0.022}$ & 0.91(43) & $0.84_{-0.03}^{+0.03}$ & $10.9_{-0.6}^{+0.7}$ & $0.521_{-0.024}^{+0.024}$ & $0.30_{-0.30}^{+0.31}$ & 0.91(42) & $0.83_{-0.02}^{+0.02}$ & $11.3_{-0.6}^{+0.6}$ & $0.528_{-0.023}^{+0.023}$ & $77.4_{-9.6}^{+9.5}$ & 0.92(42) & $3.3\times10^{-1}$ & $4.6\times10^{-1}$ \\
$4.5\pm0.5$    & $0.86_{-0.03}^{+0.03}$ & $10.0_{-0.4}^{+0.5}$ & $0.468_{-0.021}^{+0.021}$ & 1.38(42) & $0.92_{-0.03}^{+0.04}$ & $8.1_{-0.6}^{+0.6}$ & $0.425_{-0.024}^{+0.025}$ & $1.26_{-0.31}^{+0.30}$ & 1.01(41) & $0.89_{-0.03}^{+0.03}$ & $8.9_{-0.5}^{+0.6}$ & $0.433_{-0.023}^{+0.024}$ & $104.2_{-7.2}^{+7.1}$ & 1.01(41) & $2.3\times10^{-4}$ & $2.2\times10^{-4}$ \\
$5.5\pm0.5$    & $0.81_{-0.03}^{+0.03}$ & $10.5_{-0.5}^{+0.5}$ & $0.407_{-0.019}^{+0.019}$ & 1.17(39) & $0.88_{-0.04}^{+0.04}$ & $8.4_{-0.6}^{+0.7}$ & $0.364_{-0.023}^{+0.023}$ & $1.25_{-0.31}^{+0.30}$ & 0.78(38) & $0.84_{-0.03}^{+0.03}$ & $9.2_{-0.6}^{+0.7}$ & $0.374_{-0.021}^{+0.022}$ & $103.1_{-7.2}^{+7.1}$ & 0.79(38) & $5.5\times10^{-5}$ & $7.0\times10^{-5}$ \\
$7.0\pm1.0$    & $0.81_{-0.02}^{+0.02}$ & $8.9_{-0.3}^{+0.4}$ & $0.290_{-0.011}^{+0.011}$ & 1.39(48) & $0.86_{-0.03}^{+0.03}$ & $7.3_{-0.5}^{+0.5}$ & $0.260_{-0.014}^{+0.014}$ & $0.84_{-0.19}^{+0.19}$ & 1.02(47) & $0.83_{-0.02}^{+0.03}$ & $8.1_{-0.4}^{+0.5}$ & $0.268_{-0.013}^{+0.013}$ & $93.6_{-4.9}^{+4.9}$ & 1.00(47) & $9.2\times10^{-5}$ & $5.7\times10^{-5}$ \\
$9.0\pm1.0$    & $0.69_{-0.02}^{+0.02}$ & $9.7_{-0.5}^{+0.5}$ & $0.185_{-0.008}^{+0.008}$ & 1.40(41) & $0.72_{-0.03}^{+0.03}$ & $8.1_{-0.6}^{+0.7}$ & $0.158_{-0.011}^{+0.011}$ & $0.67_{-0.19}^{+0.18}$ & 1.12(40) & $0.69_{-0.02}^{+0.02}$ & $9.2_{-0.6}^{+0.6}$ & $0.170_{-0.010}^{+0.010}$ & $87.6_{-4.9}^{+4.9}$ & 1.14(40) & $1.6\times10^{-3}$ & $2.2\times10^{-3}$ \\
$11.0\pm1.0$   & $0.69_{-0.03}^{+0.03}$ & $8.7_{-0.5}^{+0.6}$ & $0.145_{-0.007}^{+0.007}$ & 1.09(38) & $0.77_{-0.04}^{+0.04}$ & $6.1_{-0.7}^{+0.8}$ & $0.112_{-0.010}^{+0.010}$ & $0.89_{-0.19}^{+0.18}$ & 0.52(37) & $0.71_{-0.03}^{+0.03}$ & $7.5_{-0.6}^{+0.7}$ & $0.125_{-0.009}^{+0.009}$ & $93.2_{-4.7}^{+4.7}$ & 0.54(37) & $1.5\times10^{-7}$ & $2.8\times10^{-7}$ \\
$14.0\pm2.0$   & $0.61_{-0.02}^{+0.02}$ & $8.3_{-0.5}^{+0.6}$ & $0.080_{-0.004}^{+0.004}$ & 1.08(42) & $0.66_{-0.03}^{+0.04}$ & $6.1_{-0.7}^{+0.8}$ & $0.061_{-0.006}^{+0.006}$ & $0.52_{-0.12}^{+0.12}$ & 0.68(41) & $0.61_{-0.02}^{+0.03}$ & $7.7_{-0.6}^{+0.7}$ & $0.073_{-0.005}^{+0.005}$ & $83.3_{-3.4}^{+3.4}$ & 0.74(41) & $9.0\times10^{-6}$ & $5.6\times10^{-5}$ \\
$19.0\pm3.0$   & $0.53_{-0.02}^{+0.02}$ & $8.1_{-0.6}^{+0.7}$ & $0.046_{-0.002}^{+0.002}$ & 1.01(42) & $0.56_{-0.03}^{+0.03}$ & $6.5_{-0.8}^{+0.9}$ & $0.035_{-0.004}^{+0.004}$ & $0.28_{-0.10}^{+0.10}$ & 0.82(41) & $0.53_{-0.02}^{+0.02}$ & $8.3_{-0.6}^{+0.7}$ & $0.047_{-0.004}^{+0.004}$ & $75.5_{-2.8}^{+2.8}$ & 0.83(41) & $2.2\times10^{-3}$ & $2.8\times10^{-3}$ \\
$25.5\pm3.5$   & $0.47_{-0.02}^{+0.02}$ & $9.1_{-0.7}^{+0.9}$ & $0.035_{-0.002}^{+0.002}$ & 1.16(42) & $0.48_{-0.03}^{+0.03}$ & $7.5_{-1.0}^{+1.1}$ & $0.026_{-0.004}^{+0.004}$ & $0.25_{-0.09}^{+0.09}$ & 1.01(41) & $0.48_{-0.02}^{+0.02}$ & $9.2_{-0.7}^{+0.8}$ & $0.038_{-0.003}^{+0.003}$ & $74.2_{-2.6}^{+2.6}$ & 0.99(41) & $8.9\times10^{-3}$ & $6.0\times10^{-3}$ \\
$36.0\pm7.0$   & $0.46_{-0.02}^{+0.02}$ & $6.6_{-0.6}^{+0.7}$ & $0.018_{-0.001}^{+0.001}$ & 0.78(48) & $0.47_{-0.02}^{+0.03}$ & $6.3_{-0.7}^{+0.9}$ & $0.017_{-0.002}^{+0.002}$ & $0.03_{-0.03}^{+0.06}$ & 0.79(47) & $0.47_{-0.01}^{+0.02}$ & $8.1_{-0.5}^{+0.6}$ & $0.027_{-0.002}^{+0.002}$ & $67.4_{-1.8}^{+1.8}$ & 0.87(47) & $5.1\times10^{-1}$ & -- \\
$50.5\pm7.5$   & $0.44_{-0.03}^{+0.03}$ & $6.9_{-0.7}^{+0.9}$ & $0.016_{-0.001}^{+0.001}$ & 1.02(50) & $0.45_{-0.04}^{+0.04}$ & $5.3_{-0.9}^{+1.2}$ & $0.010_{-0.002}^{+0.002}$ & $0.16_{-0.06}^{+0.06}$ & 0.87(49) & $0.46_{-0.02}^{+0.02}$ & $7.4_{-0.6}^{+0.7}$ & $0.021_{-0.002}^{+0.002}$ & $71.8_{-1.7}^{+1.7}$ & 0.99(49) & $4.0\times10^{-3}$ & $1.6\times10^{-1}$ \\
$66.5\pm8.5$   & $0.53_{-0.03}^{+0.03}$ & $4.6_{-0.4}^{+0.5}$ & $0.014_{-0.001}^{+0.001}$ & 0.99(52) & $0.54_{-0.04}^{+0.04}$ & $3.8_{-0.6}^{+0.7}$ & $0.011_{-0.002}^{+0.002}$ & $0.09_{-0.05}^{+0.05}$ & 0.94(51) & $0.51_{-0.02}^{+0.02}$ & $6.0_{-0.5}^{+0.5}$ & $0.021_{-0.002}^{+0.002}$ & $70.1_{-1.5}^{+1.5}$ & 1.17(51) & $5.4\times10^{-2}$ & -- \\
$90.5\pm15.5$  & $0.42_{-0.03}^{+0.03}$ & $5.4_{-0.6}^{+0.8}$ & $0.008_{-0.001}^{+0.001}$ & 0.76(59) & $0.42_{-0.03}^{+0.03}$ & $5.3_{-0.8}^{+0.8}$ & $0.008_{-0.002}^{+0.001}$ & $0.01_{-0.01}^{+0.04}$ & 0.77(58) & $0.45_{-0.01}^{+0.01}$ & $7.1_{-0.5}^{+0.5}$ & $0.018_{-0.001}^{+0.001}$ & $66.8_{-1.2}^{+1.2}$ & 0.94(58) & $8.2\times10^{-1}$ & -- \\
$119.0\pm13.0$ & $0.42_{-0.03}^{+0.03}$ & $5.7_{-0.7}^{+0.9}$ & $0.009_{-0.001}^{+0.001}$ & 0.95(57) & $0.42_{-0.04}^{+0.04}$ & $4.7_{-0.9}^{+1.2}$ & $0.006_{-0.002}^{+0.002}$ & $0.07_{-0.04}^{+0.04}$ & 0.91(56) & $0.45_{-0.02}^{+0.02}$ & $6.9_{-0.5}^{+0.6}$ & $0.017_{-0.001}^{+0.001}$ & $68.7_{-1.3}^{+1.3}$ & 1.05(56) & $8.3\times10^{-2}$ & -- \\
$145.0\pm13.0$ & $0.45_{-0.04}^{+0.04}$ & $4.9_{-0.6}^{+0.8}$ & $0.009_{-0.001}^{+0.001}$ & 1.01(57) & $0.46_{-0.05}^{+0.06}$ & $3.7_{-0.8}^{+1.2}$ & $0.005_{-0.002}^{+0.002}$ & $0.09_{-0.04}^{+0.04}$ & 0.94(56) & $0.47_{-0.02}^{+0.02}$ & $6.2_{-0.5}^{+0.5}$ & $0.016_{-0.001}^{+0.001}$ & $69.5_{-1.3}^{+1.3}$ & 1.08(56) & $3.3\times10^{-2}$ & -- \\
$176.5\pm18.5$ & $0.40_{-0.03}^{+0.04}$ & $5.1_{-0.7}^{+1.0}$ & $0.006_{-0.001}^{+0.001}$ & 1.12(65) & $0.40_{-0.04}^{+0.05}$ & $4.6_{-0.9}^{+1.3}$ & $0.005_{-0.001}^{+0.001}$ & $0.04_{-0.03}^{+0.03}$ & 1.12(64) & $0.45_{-0.01}^{+0.02}$ & $6.5_{-0.5}^{+0.5}$ & $0.015_{-0.001}^{+0.001}$ & $68.2_{-1.0}^{+1.0}$ & 1.40(64) & $3.2\times10^{-1}$ & -- \\
$207.5\pm12.5$ & $0.45_{-0.04}^{+0.04}$ & $5.0_{-0.7}^{+0.9}$ & $0.009_{-0.001}^{+0.001}$ & 0.96(57) & $0.46_{-0.06}^{+0.08}$ & $3.2_{-0.9}^{+1.4}$ & $0.004_{-0.002}^{+0.002}$ & $0.13_{-0.04}^{+0.04}$ & 0.82(56) & $0.47_{-0.02}^{+0.02}$ & $6.0_{-0.5}^{+0.6}$ & $0.015_{-0.001}^{+0.001}$ & $70.8_{-1.3}^{+1.3}$ & 0.90(56) & $1.7\times10^{-3}$ & $2.8\times10^{-2}$ \\
$234.5\pm14.5$ & $0.44_{-0.03}^{+0.04}$ & $4.9_{-0.6}^{+0.8}$ & $0.008_{-0.001}^{+0.001}$ & 1.18(59) & $0.45_{-0.05}^{+0.06}$ & $3.6_{-0.9}^{+1.2}$ & $0.005_{-0.002}^{+0.002}$ & $0.09_{-0.04}^{+0.04}$ & 1.11(58) & $0.46_{-0.02}^{+0.02}$ & $6.2_{-0.5}^{+0.5}$ & $0.015_{-0.001}^{+0.001}$ & $69.3_{-1.2}^{+1.2}$ & 1.26(58) & $3.2\times10^{-2}$ & -- \\
$263.5\pm14.5$ & $0.42_{-0.03}^{+0.04}$ & $5.3_{-0.7}^{+0.9}$ & $0.008_{-0.001}^{+0.001}$ & 1.03(59) & $0.42_{-0.05}^{+0.06}$ & $3.8_{-1.0}^{+1.4}$ & $0.004_{-0.002}^{+0.002}$ & $0.10_{-0.04}^{+0.04}$ & 0.94(58) & $0.45_{-0.02}^{+0.02}$ & $6.4_{-0.5}^{+0.5}$ & $0.015_{-0.001}^{+0.001}$ & $69.6_{-1.2}^{+1.2}$ & 1.03(58) & $1.4\times10^{-2}$ & $3.9\times10^{-1}$ \\
$291.5\pm13.5$ & $0.45_{-0.03}^{+0.03}$ & $4.9_{-0.6}^{+0.8}$ & $0.008_{-0.001}^{+0.001}$ & 0.95(57) & $0.45_{-0.04}^{+0.05}$ & $4.1_{-0.8}^{+1.0}$ & $0.006_{-0.002}^{+0.002}$ & $0.06_{-0.04}^{+0.04}$ & 0.92(56) & $0.46_{-0.02}^{+0.02}$ & $6.4_{-0.5}^{+0.5}$ & $0.016_{-0.001}^{+0.001}$ & $68.6_{-1.2}^{+1.2}$ & 1.05(56) & $1.2\times10^{-1}$ & -- \\
$327.5\pm22.5$ & $0.42_{-0.05}^{+0.05}$ & $4.4_{-0.8}^{+1.2}$ & $0.005_{-0.001}^{+0.001}$ & 0.84(69) & $0.42_{-0.09}^{+0.12}$ & $2.8_{-1.1}^{+2.0}$ & $0.002_{-0.001}^{+0.001}$ & $0.08_{-0.03}^{+0.03}$ & 0.75(68) & $0.46_{-0.02}^{+0.02}$ & $5.7_{-0.4}^{+0.5}$ & $0.012_{-0.001}^{+0.001}$ & $69.5_{-0.9}^{+0.9}$ & 1.04(68) & $3.5\times10^{-3}$ & -- \\
$369.5\pm19.5$ & $0.39_{-0.03}^{+0.04}$ & $5.5_{-0.8}^{+1.1}$ & $0.006_{-0.001}^{+0.001}$ & 1.10(65) & $0.38_{-0.04}^{+0.05}$ & $4.7_{-1.0}^{+1.5}$ & $0.004_{-0.001}^{+0.001}$ & $0.05_{-0.03}^{+0.03}$ & 1.08(64) & $0.44_{-0.01}^{+0.02}$ & $6.6_{-0.5}^{+0.5}$ & $0.014_{-0.001}^{+0.001}$ & $68.2_{-1.0}^{+1.0}$ & 1.12(64) & $1.5\times10^{-1}$ & -- \\
$411.5\pm22.5$ & $0.38_{-0.03}^{+0.04}$ & $5.3_{-0.8}^{+1.2}$ & $0.005_{-0.000}^{+0.000}$ & 0.94(68) & $0.37_{-0.05}^{+0.05}$ & $4.5_{-1.1}^{+1.5}$ & $0.003_{-0.001}^{+0.001}$ & $0.05_{-0.03}^{+0.03}$ & 0.92(67) & $0.44_{-0.01}^{+0.01}$ & $6.4_{-0.4}^{+0.5}$ & $0.013_{-0.001}^{+0.001}$ & $68.2_{-1.0}^{+1.0}$ & 1.05(67) & $1.2\times10^{-1}$ & -- \\
$455.0\pm21.0$ & $0.45_{-0.04}^{+0.04}$ & $3.9_{-0.6}^{+0.8}$ & $0.006_{-0.001}^{+0.001}$ & 0.92(68) & $0.46_{-0.05}^{+0.06}$ & $3.2_{-0.7}^{+1.0}$ & $0.004_{-0.001}^{+0.001}$ & $0.05_{-0.03}^{+0.03}$ & 0.90(67) & $0.46_{-0.01}^{+0.02}$ & $5.9_{-0.4}^{+0.4}$ & $0.014_{-0.001}^{+0.001}$ & $68.3_{-1.0}^{+1.0}$ & 1.20(67) & $1.4\times10^{-1}$ & -- \\
\hline
\end{tabular}

\begin{list}{}{}
\item[a] $R_{\rm bb}$ is estimated at a distance of 3.5 kpc. $F_{\rm bb}$ is in units of $10^{-8}~{\rm erg\,cm^{-2}\,s^{-1}}$. $R_{\rm disk}$ is the inner disk radius estimated at a distance of 3.5 kpc and an inclination angle of $\theta=60^\circ$ \citep{2019MNRAS.483..767D}. The quoted $\chi_\nu^2$ values are the reduced chi-square with degrees of freedom in parentheses. The symbol ``--'' indicates that the F-test result is invalid or unavailable.
\end{list}
\end{sidewaystable*}

\end{document}